\documentclass[12pt]{article}
\begin{document}
\title
{A new interpretation \\ of one CPT violation test\\
for $K_{0} - {\overline{K}}_{0}$ system.}
\author{K. Urbanowski\footnote{e--mail:
K.Urbanowski@if.uz.zgora.pl; K.Urbanowski@proton.if.uz.zgora.pl}
 \\ University of Zielona Gora, Institute of Physics,\\ ul. Podgorna
50, 65--246 Zielona Gora, Poland.} \maketitle

\begin{abstract}
Using a more accurate approximation than that  applied by
Lee--Oehme--Yang we show that the interpretation of the tests
measuring the difference between the $K_{0}$ mass and the
${\overline K}_{0}$ mass as the CPT--symmetry test is wrong. We
find that in fact such tests should rather be considered as tests
for the existence of the hypothetical interaction allowing the
first order $|\Delta S| = 2$ transitions $K_{0} \rightleftharpoons
{\overline K}_{0}$.
\end{abstract}

\section{Introduction}

CPT symmetry is a fundamental theorem of axiomatic quantum field
theory which follows from locality, Lorentz invariance, and
unitarity \cite{cpt}. Many tests of CPT--invariance consist in
searching for decay of neutral kaons. The proper interpretation of
them is crucial. All known CP--  and hypothetically possible
CPT--violation effects in neutral kaon complex are described by
solving the Schr\"{o}dinger--like evolution equation \cite{LOY}
--- \cite{chiu} (we use $\hbar = c = 1$ units)
\begin{equation}
i \frac{\partial}{\partial t} |\psi ; t >_{\parallel} =
H_{\parallel} |\psi ; t >_{\parallel} \label{l1}
\end{equation}
for $|\psi ; t >_{\parallel}$ belonging to the subspace ${\cal
H}_{\parallel} \subset {\cal H}$ (where ${\cal H}$ is the state
space of the physical system under investigation), e.g., spanned
by orthonormal neutral  kaons states $|K_{0}> \stackrel{\rm
def}{=} |{\bf 1}>, \; |{\overline{K}}_{0}> \stackrel{\rm def}{=}
|{\bf 2}> $, and so on, (then states corresponding to the decay
products belong to ${\cal H} \ominus {\cal H}_{\parallel}
\stackrel{\rm def}{=} {\cal H}_{\perp}$). The  nonhermitian
effective Hamiltonian $H_{\parallel}$ acts in ${\cal H}_{||}$ and

\begin{equation}
H_{\parallel} \equiv M - \frac{i}{2} \Gamma, \label{new1}
\end{equation}
where $ M = M^{+}, \;  \Gamma = {\Gamma}^{+}$, are $(2 \times 2)$
matrices.

Relations between matrix elements of  $H_{\parallel}$  implied  by
CPT--transformation properties of the Hamiltonian $H$ of the total
system, containing  neutral  kaon  complex  as  a  subsystem,  are
crucial for designing CPT--invariance and CP--violation tests  and
for the proper interpretation of their results.

The eigenstates of  $H_{\parallel}$,  $|K_{l}>$  and  $|K_{s}>$,
for the eigenvalues ${\mu}_{l}$ and ${\mu}_{s}$ respectively
\cite{LOY}
--- \cite{chiu}, \cite{ur92} --- \cite{Bigi}
\begin{equation}
{\mu}_{l(s)} = h_{0} - (+) h \equiv m_{l(s)} - \frac{i}{2}
{\gamma}_{l(s)},  \label{r5}
\end{equation}
where $m_{l(s)}, {\gamma}_{l(s)}$ are real, and
\begin{eqnarray}
h_{0} & = & \frac{1}{2}(h_{11} + h_{22}), \label{r5a} \\
h & \equiv & \sqrt{ h_{z}^{2} + h_{12} h_{21} }, \label{r5b} \\
h_{z} & = & \frac{1}{2} (h_{11} - h_{22}), \label{r5c} \\
h_{jk} & = & <{\bf j}|H_{\parallel}|{\bf k}>, \; (j,k=1,2),
\end{eqnarray}
correspond to the long (the vector $|K_{l}>$) and  short (the
vector $|K_{s}>$) living superpositions of $K_{0}$ and
$\overline{K_{0}}$.

Using the eigenvectors
\begin{equation}
|K_{1(2)}> \stackrel{\rm def}{=} 2^{-1/2} (|{\bf 1}> + (-) |{\bf
2}>) , \label{K-1,2}
\end{equation}
of the CP--transformation for the eigenvalues $\pm  1$ (we define
${\cal C}{\cal P} |{\bf 1}> = - |{\bf 2}>$, \linebreak ${\cal
C}{\cal P} | {\bf 2}> = - |{\bf 1}>$), vectors  $|K_{l}>$ and
$|K_{s}>$ can be expressed as follows
\cite{Comins,LOY-etc,{ur97-hep-ph},dafne}

\begin{equation}
|K_{l(s)}> \equiv (1 + |{\varepsilon}_{l(s)}|^{2})^{- 1/2}
[|K_{2(1)}
> + {\varepsilon}_{l(s)} |K_{1(2)} > ] , \label{K-ls}
\end{equation}
where
\begin{eqnarray}
{\varepsilon}_{l} & = & \frac{h_{12} - h_{11} + {\mu}_{l}}{h_{12}
+ h_{11} - {\mu}_{l}} \equiv - \frac{h_{21} - h_{22} +
{\mu}_{l}}{h_{21} + h_{22} - {\mu}_{l}},
\label{e-l} \\
{\varepsilon}_{s} & = & \frac{h_{12} + h_{11} - {\mu}_{s}}{h_{12}
- h_{11} + {\mu}_{l}} \equiv - \frac{h_{21} + h_{22} -
{\mu}_{s}}{h_{21} - h_{22} + {\mu}_{s}}, \label{e-s}
\end{eqnarray}
This form of $|K_{l}>$ and $|K_{s}>$ is used in many papers when
possible departures from CP-- or CPT--symmetry in  the system
considered are discussed.

One can easily  verify that if ${\mu}_{l} \neq {\mu}_{s} $ then:
\begin{equation}
h_{11} = h_{22} \; \Longleftrightarrow \;{\varepsilon}_{l} =
{\varepsilon}_{s}. \label{e-l=s}
\end{equation}

The experimentally measured values of the parameters
${\varepsilon}_{l}, {\varepsilon}_{s}$ are very small for neutral
kaons. Assuming
\begin{equation}
|{\varepsilon}_{l}| \ll 1, \; \; |{\varepsilon}_{s}| \ll 1,
\label{e<1}
\end{equation}
one can, e.g. find:
\begin{equation}
h_{11} - h_{22} \simeq ({\mu}_{s} - {\mu}_{l}) ({\varepsilon}_{s}
- {\varepsilon}_{l} ). \label{h-h-simeq}
\end{equation}

Keeping in mind that $h_{jk} = M_{jk} - \frac{i}{2} {\Gamma}_{jk},
M_{kj} = M_{jk}^{\ast},{\Gamma}_{kj} = {\Gamma}_{jk}^{\ast}$
(where $M_{jj} \stackrel{\rm def}{=} M_{j}, \; (j =1,2)$) and
using (\ref{h-h-simeq}) one can find, among others that
\cite{ur97-hep-ph,dafne}
\begin{eqnarray}
\Re \, (h_{11} - h_{22})  \equiv M_{1} - M_{2} & \simeq & -
({\gamma}_{s} - {\gamma}_{l}) [ \tan {\phi}_{SW}\,  \Re \,
(\frac{{\varepsilon}_{s}
 - {\varepsilon}_{l}}{2})  \nonumber \\
& \; & - \Im \,(\frac{{\varepsilon}_{s} - {\varepsilon}_{l}}{2})
\, ], \label{test}
\end{eqnarray}
where $\Re \, (z)$ denotes the real part, $\Im \, (z)$ - the
imaginary part of a complex number $(z)$, and
\begin{equation}
\tan {\phi}_{SW} \stackrel{\rm def}{=} \frac{2(m_{l} -
m_{s})}{{\gamma}_{s} - {\gamma}_{l}} . \label{tan-phi}
\end{equation}
Usually it is assumed that $M_{1}, M_{2}$ denote the masses of the
particle "1" and its antiparticle "2" \cite{LOY}--- \cite{data}).

One should remember that relations (\ref{h-h-simeq}), (\ref{test})
are valid only if condition (\ref{e<1}) holds. On the other hand
it is appropriate to emphasize at this point that all relations
(\ref{K-ls}) --- (\ref{test}) do not depend on the specific form
of the effective Hamiltonian $H_{\parallel}$ and that they do not
depend on the approximation used to calculate $H_{||}$ . They are
induced by geometric relations between various base vectors in
two--dimensional subspace ${\cal H}_{\parallel}$ but the
interpretation of the relation, eg., (\ref{test}), depends on
properties of the matrix elements $h_{jk}$ of the effective
Hamiltonian $H_{\parallel}$. This means that, if for example,
$H_{\parallel} \neq H_{LOY}$, where $H_{LOY}$ is the
Lee--Oehme--Yang effective Hamiltonian, then the interpretation of
the relation (\ref{test}), e.g., as the CPT symmetry test, and
other similar relations need not be the same for $H_{\parallel}$
and for $H_{LOY}$.

The aim of this note is to analyze the interpretation of the test
based on the relation (\ref{test}) which is commonly considered as
the CPT violation test \cite{Comins} --- \cite{LOY-etc},
\cite{ur97-hep-ph,dafne,data,Lee-qft}. Such an interpretation
follows from the properties of the matrix elements of $H_{LOY}$.

In the kaon rest frame, the time evolution is governed by the
Schr\"{o}dinger equation
\begin{equation}
i \frac{\partial}{\partial t} U(t)|\phi > = H U(t)|\phi >,  \; \;
U(0) = I, \label{Sr-eq}
\end{equation}
where $I$ is the unit operator in $\cal H$, $|\phi > \equiv |\phi
; t_{0} = 0> \in {\cal H}$  is  the  initial  state  of  the
system:
\begin{equation}
|\phi  >  \equiv  |\psi  >_{\parallel} \equiv q_{1}|{\bf 1}> +
q_{2}|{\bf 2}> \, \in {\cal H}_{||} \subset {\cal H} \label{init}
\end{equation}
(in  our case  $|\phi ;t> = U(t) |\phi >$), $H$ is the total
(selfadjoint) Hamiltonian, acting in $\cal H$. Thus the  total
evolution  operator $U(t)$ is unitary.

Starting from the Schr\"{o}dinger equation and using the
Weisskopf--Wigner method Lee, Oehme and Yang derived the following
formula for the matrix elements $h_{jk}^{LOY}, \, (j,k =1,2)$ of
their effective Hamiltonian $H_{LOY}$ (see
\cite{LOY,Gaillard,Comins,ur-pi-00}):
\begin{eqnarray}
h_{jk}^{LOY} & = &  m_{0}\, {\delta}_{j,k} - {\Sigma}_{jk} (m_{0}
)   \label{h-jk-LOY} \\
& = & M_{jk}^{LOY} - \frac{i}{2} {\Gamma}_{jk}^{LOY}, \; \; \;
(j,k = 1,2) , \label{M-jk-LOY}
\end{eqnarray}
where ${\Sigma}_{jk}(\epsilon) = <{\bf j}|\Sigma (\epsilon)|{\bf
k}>, \, (j,k = 1,2)$, and
\begin{equation}
\Sigma ( \epsilon ) = PHQ \frac{1}{QHQ - \epsilon - i 0} QHP =
{\Sigma}^{R}(\epsilon) + i {\Sigma}^{I}(\epsilon), \label{Sigma}
\end{equation}
and ${\Sigma}^{R}(\epsilon = {\epsilon}^{\ast}) =
{\Sigma}^{R}(\epsilon = {\epsilon}^{\ast})^{+}, \,
{\Sigma}^{I}(\epsilon = {\epsilon}^{\ast}) = {\Sigma}^{I}(\epsilon
= {\epsilon}^{\ast})^{+}$, $P$ is the projector operator onto the
subspace ${\cal H}_{||}$:
\begin{equation}
P \equiv |{\bf 1}><{\bf 1}| + |{\bf 2}><{\bf 2}|, \label{P}
\end{equation}
$Q$ is the projection operator onto the subspace of decay products
${\cal H}_{\perp}$:
\begin{equation}
Q \equiv I - P, \label{Q}
\end{equation}
and vectors $|{\bf 1}>$, $|{\bf 2}>$ considered above are the
eigenstates of the free Hamiltonian, $H^{(0)}$, ( here $H \equiv
H^{(0)} + H^{I}$), for 2-fold degenerate eigenvalue $m_{0}$:
\begin{equation}
H^{(0)} |{\bf j} > = m_{0} |{\bf j }>, \; \;  (j = 1,2),
\label{H-0-j}
\end{equation}
$H^{I}$ denotes the interaction which is responsible for the decay
process. This means that
\begin{equation}
[P, H^{(0)}] = 0. \label{P-H-0}
\end{equation}
The condition guaranteeing the occurrence of transitions  between
subspaces ${\cal H}_{\parallel}$ and ${\cal H}_{\perp}$, i.e., the
decay process of states in ${\cal H}_{\parallel}$, can be written
as follows
\begin{equation}
[P,H^{I}] \neq 0, \label{P-H-I-neq-0}
\end{equation}
that is
\begin{equation}
[P,H] \neq 0, \label{[P,H]}
\end{equation}

The operator $H_{LOY}$ has the following form
\begin{equation}
H_{LOY} = PHP  - \Sigma (m_{0}) \equiv PHP -  \Sigma (m_{0}),
\label{H-LOY}
\end{equation}
where $PHP \equiv PH^{(0)}P + PH^{I}P$, and, in the considered
case
\begin{equation}
PH^{(0)}P \equiv m_{0} P, \label{LOY=m0P}
\end{equation}
and (see \cite{LOY,Gaillard,Comins})
\begin{equation}
PH^{I}P \equiv 0. \label{LOY=ds=1-3}
\end{equation}

Following the method used by LOY, assumption (\ref{LOY=ds=1-3})
may be either kept or dropped. This has no effect on the CP-- and
CPT--transformation properties of the matrix elements of $H_{LOY}$
and conclusions derived in this paper. The assumption
(\ref{LOY=ds=1-3}) is a reflection of the opinion of physicists
following the ideas of \cite{LOY} and deriving and then applying
$H_{LOY}$ that matrix elements of $H^{I}$ are too small in
comparison with $m_{0}$ in order to have any measurable effect on
time evolution in ${\cal H}_{||}$ (see
\cite{LOY,Gaillard,Comins}). Indeed, in some papers (see, eg.,
\cite{LOY1,LOY-etc,Cheng,Lee-qft}), instead of (\ref{LOY=ds=1-3})
the assumption that $PH^{I}P \neq 0$ is introduced by hand into
formulae for matrix elements of $H_{LOY}$ without verifying
whether it results in the final form of $H_{LOY}$ or not. It
appears that the properties of such $H_{LOY}$ do not differ from
the properties of the LOY effective Hamiltonian derived within the
use of the assumption (\ref{LOY=ds=1-3}). In other words, the
approximation applied by LOY in \cite{LOY} leads to the operator
$H_{LOY}$ whose CP-- and CPT--transformation properties do not
depend on whether $PH^{I}P \neq 0$, that is $H_{12} \stackrel{\rm
def}{=} <{\bf 1}|H|{\bf 2}> \neq 0$, or not. (Note that within the
LOY assumptions  $H_{12} \equiv <{\bf 1}|H^{I}|{\bf 2}>$). This
property of the LOY approximation means that, for example, the
verification of the presence (or absence) of the hypothetical
superweak interactions causing the direct, first order $K_{0}
\rightleftharpoons {\overline{K}}_{0}$, $| \Delta S |= 2$,
transitions \cite{Wolfenstein,Jarlskog} in experiments designed
within use of the LOY theory is very difficult and that the
interpretation of results of such experiments can not be
considered as definitive. The same refers to the problem of how to
identify within the LOY theory effects caused by such hypothetical
interactions and similar effects predicted by the Standard Model
(SM) \cite{Isidori,Bigi,Buras}. In this place, in order to avoid
possible misunderstandings, one should explain that from the point
of view of the problems discussed in this paper, SM effective
Hamiltonians, $H_{eff}$, for the problem under consideration,
should be identified with $H_{||}$ (or $H_{LOY}$) but they can not
be identified with the operator $PHP$.

Usually, in LOY and related approaches, it is assumed that
\begin{equation}
{\Theta}H^{(0)}{\Theta}^{-1} = {H^{(0)}}^{+} \equiv H^{(0)} ,
\label{Theta-H-0}
\end{equation}
where $\Theta$ is the antiunitary operator
\cite{messiah,bohm,Wigner}:
\begin{equation}
\Theta \stackrel{\rm def}{=} {\cal C}{\cal P}{\cal T}.
\label{Theta=CPT}
\end{equation}

The subspace of neutral kaons ${\cal H}_{\parallel}$ is assumed to
be invariant under $\Theta$:
\begin{equation}
{\Theta} P {\Theta}^{-1} = P^{+} \equiv P. \label{Theta-P}
\end{equation}

Now, if in addition to (\ref{Theta-H-0})
${\Theta}H^{I}{\Theta}^{-1} = H^{I}$, that is if
\begin{equation}
[ \Theta , H] = 0, \label{CPT-H}
\end{equation}
then using, e.g., the following phase convention
\begin{equation}
\Theta |{\bf 1}> \stackrel{\rm def}{=} e^{{\textstyle
i\alpha}_{\mit\Theta}} |{\bf 2}>, \;\; \Theta|{\bf 2}>
\stackrel{\rm def}{=} e^{{\textstyle i\alpha}_{\mit\Theta}} |{\bf
1}>, \label{cpt1}
\end{equation}
and taking into account that $< \psi | \varphi > =
<{\Theta}{\varphi}|{\Theta}{\psi}>$, one easily finds from
(\ref{h-jk-LOY}), (\ref{Sigma})  that
\begin{equation}
h_{11}^{LOY} - h_{22}^{LOY} = 0  \label{LOYh11=h22}
\end{equation}
in the CPT--invariant system. This  is  the standard result of the
LOY approach and this is the picture  which one meets in  the
literature \cite{LOY}  ---  \cite{chiu}, \cite{dafne,Lee-qft}.
Property (\ref{LOYh11=h22}) leads to the conclusion that (see
(\ref{e-l=s}))
\begin{equation}
{\varepsilon}_{l} - {\varepsilon}_{s} = 0. \label{LOY-e-l=s}
\end{equation}
Therefore the tests based on the relation (\ref{test}) are
considered as the test of CPT--invariance and the results of such
tests are interpreted that the masses of the particle "1" (the
$K_{0}$ meson) and its antiparticle "2" (the ${\overline{K}}_{0}$
meson) must be equal if CPT--symmetry holds. Parameters appearing
in the right side of the relation (\ref{test}) can be extracted
from experiments in such tests and then these parameters can be
used to estimate the left side of this relation. The estimation
for the mass difference obtained in this way with the use of the
recent data \cite{data} reads
\begin{equation}
\frac{|M_{1} - M_{2}|}{m_{K_{0}}} = \frac{|m_{K_{0}} -
m_{{\overline{K}}_{0}}|}{m_{K_{0}}} \leq 10^{-18}, \label{mk-mk}
\end{equation}
and this estimation is considered as indicating no
 CPT--violation effect. This interpretation follows
from the properties of the $H_{LOY}$.

Note that in fact the above interpretation of the relation
(\ref{mk-mk}) could be considered as the confirmation of the CPT
invariance only if the property (\ref{LOYh11=h22}) were the exact
relation. The accuracy of the LOY approximation was discussed,
eg., in \cite{chiu,leonid1,leonid2,nowakowski,wang}. According to
these and other papers one can determine all parameters needed for
the time evolution formulae in LOY theory up to the accuracy of
$\frac{{\Gamma}_{X}}{m_{X}} \sim 10^{-15}$, where $X=K_{s},K_{l}$,
in terms of known quantities. In Khalfin's papers
\cite{leonid1,leonid2}, where the exact theory of time evolution
in neutral $K$ complex was discussed, one meets an opinion that
the magnitude of some "new effects" predicted within such an
approach can even be of the order $\frac{{\Gamma}_{l}}{m_{s} -
m_{l}} \sim 10^{-3}$. In \cite{chiu,nowakowski} the same Khalfin's
"new effect" was estimated to be of order $<10^{-11}$. All these
estimations and obtained in \cite{ur02-hep-ph} rigorous result
that $h_{11}$ can not be equal to $h_{22}$ in CPT invariant and CP
noninvariant system show that the interpretation of test for
neutral $K$ subsystem within the LOY theory can have more than one
meaning and it can not be considered as crucial. This remark
concerns especially tests of type (\ref{test}) (where the result
is of order given in (\ref{mk-mk})). So one should look for much
more accurate approximations describing the time evolution in
neutral $K$ complex and should try to interpret results of tests,
eg., of type (\ref{test}) within the use of this more accurate
approximation and of the exact theory.

\section{Beyond the LOY approximation}

The more exact approximate formulae for $H_{||}$ than those
obtained within the LOY approach (i.e. than $H_{LOY}$) can be
derived using  the  Krolikowski--Rzewuski (KR) equation for the
projection of a state vector \cite{KR,ur-pra-94}, \cite{ur92}
--- \cite{ur95},  \cite{Pi00}. KR equation results from
the Schr\"{o}dinger  equation  (\ref{Sr-eq}) for  the total system
under consideration \cite{KR} (see also, e.g., \cite{Rocky}) and
it is the exact evolution equation for the subspace ${\cal H}_{||}
\subset {\cal H}$. In the case of initial conditions of the type
(\ref{init}), KR equation takes the following form
\begin{equation}
( i \frac{\partial}{ {\partial} t} - PHP )
U_{\parallel}(t)|\psi>_{||}
 =  - i \int_{0}^{\infty} K(t - \tau ) U_{\parallel}
( \tau )|\psi>_{||} d \tau,   \label{KR1}
\end{equation}
where $U_{||}(t) \equiv PU(t)P$ is the nonunitary evolution
operator for the subspace ${\cal H}_{||}$, $U_{||}(t)|\psi
>_{||}$  $= |\psi ;t>_{||} \in {\cal H}_{||}$,  $ U_{\parallel} (0) = P$,  and
\begin{eqnarray}
K(t) & = & {\mit \Theta} (t) PHQ \exp (-itQHQ)QHP, \label{K(t)} \\
{\mit \Theta} (t) & = & { \{ } 1 \;{\rm for} \; t \geq 0, \; \; 0
\; {\rm for} \; t < 0 { \} } . \nonumber
\end{eqnarray}

The integro--differential equation (\ref{KR1}) can be replaced by
the following differential one (see \cite{KR,ur-pra-94},
\cite{ur92} --- \cite{ur95},  \cite{Pi00})
\begin{equation}
( i \frac{\partial}{ {\partial} t} - PHP - V_{||}(t) )\,
U_{\parallel}(t)|\psi>_{||} = 0. \label{KR2}
\end{equation}
This equation has the required form of the Schr\"{o}dinger--like
equation (\ref{l1}) with the effective Hamiltonian $H_{||}$, which
in general depends on time $t$ \cite{KR,horwitz},
\begin{equation}
H_{||} \equiv H_{\parallel} (t) \stackrel{\rm def}{=} PHP +
V_{\parallel}(t). \label{H||(t)}
\end{equation}
%This $H_{||}$ is very sensitive to the properties of the matrix
%elements of $PHP$.

In the case (\ref{[P,H]}), to the lowest nontrivial order, the
following formula for $V_{||}(t)$ has been found in
\cite{ur93,ur-pra-94}
\begin{equation}
V_{\parallel}(t) \cong V_{\parallel}^{(1)} (t) \stackrel{\rm
def}{=} -i \int_{0}^{\infty} K(t - \tau ) \exp {[} i ( t - \tau )
PHP {]} d \tau . \label{V||(t)}
\end{equation}
All steps leading to the approximate formula  (\ref{V||(t)}) for
$V_{\parallel}^{(1)} (t)$ are well defined (see
\cite{ur93,ur-pra-94}).

 We are rather interested in the properties of the system at long
time period, at the same for which the LOY approximation was
calculated,  and therefore we will consider the properties of
\begin{equation}
V_{||} \stackrel{\rm def}{=}\lim_{t \rightarrow \infty}
V_{||}^{(1)}(t), \label{V||-infty}
\end{equation}
instead of the general case $V_{||}(t) \cong V_{||}^{(1)}(t)$.

For simplicity we assume that the CPT--symmetry is conserved in
our system, that is that the condition (\ref{CPT-H}) holds. The
consequence of this  assumption is that
\begin{equation}
H_{11} = H_{22} \stackrel{\rm def}{=} H_{0}, \label{H_0}
\end{equation}
where
\begin{equation}
 H_{jk} = <{\bf j}|H|{\bf k}>,  \label{H-jk}
\end{equation}
and $(j,k= 1,2)$. In this case the matrix elements of $\Sigma
(\epsilon)$ have the following properties \cite{ur93,ur95,ur98}
\begin{equation}
{\Sigma}_{11} ( \epsilon = {\epsilon}^{\ast} ) \equiv
{\Sigma}_{22} ( \epsilon = {\epsilon}^{\ast} ) \stackrel{\rm
def}{=} {\Sigma}_{0} ( \epsilon = {\epsilon}^{\ast} ).
\label{sigma-jk-0}
\end{equation}

So, in the case of the projector $P$ given by the formula
(\ref{P}) for
\begin{equation}
PHP \equiv H_{0}\, P, \label{P-H12=0}
\end{equation}
that is for
\begin{equation}
H_{12} = H_{21} = 0,  \label{H12=0}
\end{equation}
one finds that
\begin{equation}
V_{||} = - \Sigma (H_{0}). \label{V-H-0}
\end{equation}
This means that in the case (\ref{P-H12=0})
\begin{equation}
H_{||} = H_{0} \, P - \,\Sigma (H_{0}), \label{H||-H12=0}
\end{equation}
(where $H_{||} \stackrel{\rm def}{=} \lim_{t \rightarrow \infty}
H_{||}(t) \equiv PHP + \lim_{t \rightarrow \infty}V_{||}(t)$),
that is exactly as in the LOY approach (see (\ref{H-LOY})). This
also means that in such a case simply $(h_{11} - h_{22}) = 0$ when
CPT symmetry is conserved.

On the other hand, in the case
\begin{equation}
H_{12} = H_{21}^{\ast} \neq 0, \label{H12n0}
\end{equation}
and conserved CPT, one obtains \cite{Pi00}
\begin{eqnarray}
V_{||} & = & - \frac{1}{2} \Sigma (H_{0} + |H_{12}|)\, \Big[ \Big(
1 - \frac{H_{0}}{|H{_{12}|}} \Big)P + \frac{1}{|H_{12}|} PHP \Big]
\nonumber \\
& &  - \frac{1}{2} \Sigma (H_{0} - |H_{12}|)\, \Big[ \Big( 1 +
\frac{H_{0}}{|H{_{12}|}} \Big)P - \frac{1}{|H_{12}|} PHP \Big].
\label{V-H12n0}
\end{eqnarray}
Matrix  elements  $v_{jk} = <{\bf j}|V_{||}|{\bf k}>, \, (j,k
=1,2)$ of this $V_{||}$ we are interested in take the following
form
\begin{eqnarray}
v_{j1} = & - & \frac{1}{2} {\Big\{ } {\Sigma}_{j1} (H_{0} + |
H_{12} |)
+ {\Sigma}_{j1} (H_{0} - | H_{12} |)  \nonumber  \\
& + & \frac{H_{21}}{|H_{12}|} {\Sigma}_{j2} (H_{0} + | H_{12} |) -
\frac{H_{21}}{|H_{12}|} {\Sigma}_{j2} (H_{0} - | H_{12} |) {\Big\}
} ,
\nonumber \\
& \; &   \label{v-jk} \\
v_{j2} = & - & \frac{1}{2} {\Big\{ } {\Sigma}_{j2} (H_{0} + |
H_{12} |)
+ {\Sigma}_{j2} (H_{0} - | H_{12} |)  \nonumber  \\
& + & \frac{H_{12}}{|H_{12}|} {\Sigma}_{j1} (H_{0} + | H_{12} |) -
\frac{H_{12}}{|H_{12}|} {\Sigma}_{j1} (H_{0} - | H_{12} |) {\Big\}
}. \nonumber
\end{eqnarray}

The form of these matrix elements is rather inconvenient for
searching for their properties depending on the matrix elements
$H_{12}$ of $PHP$. It can be done relatively simply assuming
\cite{ur95,ur98}
\begin{equation}
|H_{12}| \ll |H_{0} | . \label{H12<H0}
\end{equation}
Within such an assumption one finds \cite{ur95,ur98}
\begin{equation}
v_{j1} \simeq - {\Sigma}_{j1} (H_{0} ) - H_{21} \frac{ \partial
{\Sigma}_{j2} (x) }{\partial x}
\begin{array}[t]{l} \vline \, \\ \vline \,
{\scriptstyle x = H_{0} } \end{array} , \label{vj1-H12<}
\end{equation}
\begin{equation}
v_{j2} \simeq - {\Sigma}_{j2} (H_{0} ) - H_{12} \frac{ \partial
{\Sigma}_{j1} (x) }{\partial x}
\begin{array}[t]{l} \vline \, \\ \vline \,
{\scriptstyle x = H_{0} } \end{array} , \label{vj2-H12<}
\end{equation}
where $j = 1,2$.  One  should  stress  that  due  to  the presence
of resonance terms, derivatives $\frac{\partial}{\partial x}
{\Sigma}_{jk} (x)$ need not  be  small,  and  so  the  products
$H_{jk} \frac{\partial}{\partial x} {\Sigma}_{jk}  (x)$  in
(\ref{vj1-H12<}), (\ref{vj2-H12<}).

From this formulae we  conclude  that, e.g., the  difference
between the diagonal   matrix   elements   which plays an
important role in designing tests of type (\ref{test}) for  the
neutral kaons system, equals to the lowest order of  $|H_{12}|$,

\begin{equation}
h_{11} - h_{22} \simeq H_{12} \frac{ \partial {\Sigma}_{21} (x) }
{\partial x}
\begin{array}[t]{l} \vline \, \\ \vline \,
{\scriptstyle x = H_{0} } \end{array} - H_{21} \frac{ \partial
{\Sigma}_{12} (x) }{\partial x}
\begin{array}[t]{l} \vline \, \\ \vline \,
{\scriptstyle x = H_{0} }\end{array} \neq 0. \label{h11-h22-H12<}
\end{equation}

So, in a general case, in contradiction to the property
(\ref{LOYh11=h22}) obtained within the LOY theory, one finds for
diagonal matrix elements  of $H_{\parallel}$ calculated within the
above described approximation  that  in CPT--invariant system the
nonzero matrix elements, $H_{12} \neq 0$, of $PHP$ cause that
$(h_{11} - h_{22}) \neq 0$.

From the formula (\ref{h11-h22-H12<}) it follows that the left
side of the relation (\ref{test}) takes the following form in the
case of very weak interactions allowing for the nonzero first
order transitions $|{\bf 1}> \rightleftharpoons |{\bf 2}>$
\begin{equation}
M_{1} - M_{2} \, = \, \Re \, (h_{11} - h_{22}) = 2\, \Im \, \Big(
H_{21} { \frac{\partial {\Sigma}_{12}^{I}(x)}{\partial x} \,
\vrule \,}_{x=H_{0}} \, \Big) \, + \, \ldots  \, \neq \, 0.
\label{Re-h11-h22}
\end{equation}
(Note that as a matter of fact assuming  (\ref{H-0-j}) one has
$H_{21} \equiv <{\bf 2}|H^{I}|{\bf 1}>$ in (\ref{Re-h11-h22})).
Thus taking into account this result and the implications of the
assumptions (\ref{P-H12=0}), (\ref{H12=0}) one can conclude that
\begin{equation}
\Re \, (h_{11} - h_{22})\, = 0 \, \Leftrightarrow \, |H_{12}| \, =
\, 0 \label{Re=0}
\end{equation}
within the considered approximation. Finally, using result
(\ref{Re-h11-h22}) one can replace the relation (\ref{test}) by
the following one:
\begin{eqnarray}
2\, \Im \, \Big( <{\bf 2}|H^{I}|{\bf 1}> { \frac{\partial
{\Sigma}_{12}^{I}(x)}{\partial x} \, \vrule \,}_{x=H_{0}} \, \Big)
& \simeq & - ({\gamma}_{s} - {\gamma}_{l}) [ \tan {\phi}_{SW}\,
\Re \, (\frac{{\varepsilon}_{s}
 - {\varepsilon}_{l}}{2})   \nonumber \\
& \; & - \Im \,(\frac{{\varepsilon}_{s} - {\varepsilon}_{l}}{2})
\, ]. \label{test-a}
\end{eqnarray}

\section{Discussion}

The results (\ref{h11-h22-H12<})  and (\ref{Re-h11-h22}) are in
full agreement with the conclusions drawn in \cite{ur02-hep-ph} on
the ground of basic assumptions of quantum theory. Note that
similar relation to (\ref{h11-h22-H12<}) was obtained for CPT
invariant system in \cite{Leonid} by means of another, more
accurate than LOY, approximation. In \cite{ur02-hep-ph} it has
been shown that the diagonal matrix elements of the exact
effective Hamiltonian governing the time evolution in the subspace
of states of an unstable particle and its antiparticle can not be
equal at $t > t_{0} =0$ ($t_{0}$ is the instant of  creation of
the pair) when the total system under consideration is CPT
invariant but CP noninvariant. The proof of this property is
rigorous. The unusual consequence of this result is that in such a
case, contrary to the properties of stable particles, the masses
of the unstable particle "1" and its antiparticle "2" need not be
equal for $t \gg t_{0}$. In fact there is nothing strange in this
conclusion. From (\ref{CPT-H}) (or from the CPT theorem \cite{cpt}
of axiomatic quantum field theory) it only follows that the masses
of particle and antiparticle eigenstates of $H$ (i.e., masses of
stationary states for $H$) should be the same in the CPT invariant
system. Such a conclusion can not be drawn from (\ref{CPT-H}) for
particle "1" and its antiparticle "2" if they are unstable, ie.,
if states $|{\bf 1}>, |{\bf 2}>$ are not eigenstates of $H$. There
is no axiomatic quantum field theory of unstable particles.

In this place one should explain that the property $H_{12}=H_{21}
= 0$, which according to (\ref{h11-h22-H12<}) implies that
$(h_{11} - h_{22}) = 0$ in the considered approximation, does not
mean that the predictions following from the use of the exact
effective Hamiltonian (or the more accurate effective Hamiltonian
than the LOY theory) should lead to the the same masses for
$K_{0}$ and for ${\overline{K}}_{0}$. This does not contradict the
above mentioned conclusion  about masses of unstable particles
drawn in \cite{ur02-hep-ph} for the exact $H_{||}$: Simply, in the
case $H_{12}=H_{21} = 0$ the mass difference is very, very small
and should arise at higher orders of the more accurate
approximation described in Sec.2.

Using the above, briefly described formalism, one can find
$(h_{11} - h_{22})$ for the generalized Fridrichs--Lee model
\cite{chiu,ur93}. Within this toy model one finds \cite{ur98}
\begin{eqnarray}
\Re \, (h_{11}  -  h_{22}) \stackrel{\rm df}{=} \Re \,
(h_{11}^{FL} - h_{22}^{FL}) & \simeq & i \frac{
m_{21}{\Gamma}_{12} - m_{12}{\Gamma}_{21} }{4(m_{0} - \mu )}
\nonumber \\
& \equiv & \frac{{\Im \,}(m_{12}{\Gamma}_{21})}{2(m_{0}- \mu )}.
\label{FL1}
\end{eqnarray}
This estimation has been obtained in the case of conserved
CPT--symmetry for $|m_{12}| \ll (m_{0}- \mu)$, which corresponds
to (\ref{H12<H0}). In (\ref{FL1}) ${\Gamma}_{12}, {\Gamma}_{21}$
can be identified with those appearing in  the LOY theory, $m_{0}
\equiv H_{11} = H_{22}$ can be considered  as  the kaon mass
\cite{chiu}, $m_{jk} \equiv H_{jk} \, (j,k =1,2)$, $\mu$ can be
treated as the mass of the decay products of the neutral  kaon
\cite{chiu}.

For neutral $K$-system, to evaluate $(h_{11}^{FL} - h_{22}^{FL})$
one can follow, e.g., \cite{chiu,dafne} and one can take
$\frac{1}{2}{\Gamma}_{21} = \frac{1}{2}{\Gamma}_{12}^{\ast} \sim
\frac{1}{2}{\Gamma}_{s} \sim 5 \times 10^{10} {\rm sec}^{-1}$ and
$(m_{0} - \mu ) = m_{K} - 2m_{\pi} \sim 200$ MeV $\sim 3 \times
10^{23} {\rm sec}^{-1}$ \cite{data}. Thus
\begin{equation}
\Re \, (h_{11}  -  h_{22}) \sim \frac{{\Gamma}_{s}}{4(m_{K}-
2m_{\pi} )}\, \Im\,(H_{12}), \label{FL1a}
\end{equation}
that is,
\begin{equation}
|\Re \, (h_{11}^{FL} - h_{22}^{FL})| \sim  1,7 \times 10^{-13}
|\Im\,(m_{12})|\equiv 1,7 \times 10^{-13} |\Im\,(H_{12})| .
\label{FL2}
\end{equation}

Note that the relation (\ref{FL1a}) is equivalent to the following
one
\begin{equation}
\Re \, (h_{11}  -  h_{22}) \sim - i \frac{{\Gamma}_{s}}{4(m_{K}-
2m_{\pi} )} <{\bf 1}|H_{-}|{\bf 2}>, \label{FL1b}
\end{equation}
where $H_{-}$ is the CP odd part of the total Hamiltonian $H
\equiv H_{+} + H_{-}$. There are $H_{-} \stackrel{\rm def}{=}
\frac{1}{2}[H - ({\cal C \cal P} ) H ({\cal C \cal P} )^{+}]$ and
$H_{+} \stackrel{\rm def}{=} \frac{1}{2}[H + ({\cal C \cal P} ) H
({\cal C \cal P} )^{+}]$ (see \cite{Lee-qft,Bigi}). $H_{+}$
denotes the CP even part of $H$. We have $<{\bf 1}|H_{-}|{\bf 2}>
\equiv i  \Im \, (<{\bf 1}|H|{\bf 2}>)$\linebreak $= i\,\Im
\,(H_{12})$. According to the literature, in the case of the
superweak model for CP violation it should be $<{\bf 1}|H_{-}|{\bf
2}> \equiv i \Im\,(H_{12})$ $ \neq 0$ to the lowest order and
$<{\bf 1}|H_{-}|{\bf 2}> = 0$ in the case of a miliweak model
\cite{Lee-qft,Bigi}.

For the Fridrichs--Lee model it has been found in \cite{ur93} that
$h_{jk}(t) \simeq h_{jk}$ practically for $t \geq T_{as} \simeq
\frac{10^{2}}{\pi (m_{0} - |m_{12}| - \mu )}$. This leads to the
following estimation of $T_{as}$ for the neutral $K$--system:
$T_{as} \sim 10^{-22}$ sec.

Dividing both sides of (\ref{FL2}) by $m_{0}$ one arrives at the
relation corresponding to (\ref{mk-mk}):
\begin{equation}
\frac{|\Re \, (h_{11}^{FL} - h_{22}^{FL})|}{m_{0}} \sim  1,7
\times 10^{-13} \frac{|\Im\,(m_{12})|}{m_{0}}\equiv 1,7 \times
10^{-13} \frac{|\Im\,(H_{12})|}{m_{0}}. \label{FL3}
\end{equation}
So, if we suppose for a moment that the result (\ref{mk-mk}) is
the only experimental result for neutral $K$ complex then it is
sufficient for $\frac{|\Im\,(H_{12})|}{m_{0}}$ to be
$\frac{|\Im\,(H_{12})|}{m_{0}} < 10^{-5}$ in order to fulfil the
estimation (\ref{mk-mk}). Of course this could be considered as
the upper bound for a possible value of the ratio
$\frac{|\Im\,(H_{12})|}{m_{0}}$ only if there were no other
experiments and no other data for the $K_{0}, {\overline{K}}_{0}$
complex. Note that form such a point o view the suitable order of
$\frac{|\Im\,(H_{12})|}{m_{0}}$ is easily reached by the
hypothetical Wolfenstein superweak interactions
\cite{Wolfenstein}, which admits first order $|\Delta S| = 2$
transitions $K_{0} \rightleftharpoons {\overline K}_{0}$, that is,
which assumes a non--vanishing first order transition matrix
$H_{12} = <{\bf 1}|H^{I}|{\bf 2}> \sim g G_{F} \neq 0$ with $ g
\ll G_{F}$. The more realistic estimation for
$\frac{|\Im\,(H_{12})|}{m_{0}}$ can be found using the property
$\frac{|\Im\,(H_{12})|}{m_{0}} \equiv \frac{|<{\bf 1}|H_{-}|{\bf
2}>|}{m_{0}}$. One can assume that $\frac{|<{\bf 1}|H_{-}|{\bf
2}>|}{m_{0}} \sim \frac{H_{-}}{H_{strong}}$. There is
$\frac{H_{-}}{H_{strong}} \sim 10^{-14} |\varepsilon |$ for the
case of the hypothetical superweak interactions (see
\cite{Lee-qft}, formula (15.138)) and thus
$\frac{|\Im\,(H_{12})|}{m_{0}} \sim 10^{-14} |\varepsilon |$.
(Using this last estimation one should remember that it follows
from the LOY theory of neutral $K$ complex). This estimation
allows one to conclude that
\begin{equation}
\frac{|\Re \, (h_{11}^{FL} - h_{22}^{FL})|}{m_{0}} \sim  1,7
\times 10^{-27}|\varepsilon |. \label{FL4}
\end {equation}
This estimation is the estimation of type (\ref{mk-mk}) and it can
be considered as a lower bound for $\frac{|\Re \, (h_{11} -
h_{22})|}{m_{0}}$. (see also \cite{Piskorski2003}).

Note that contrary to the approximation described in Sec. 2, the
LOY approximation, as well as the similar approximation leading to
the Bell--Strinberger unitary relations \cite{ Bell} are unable to
detect and correctly identify effects caused by the existence (or
absence) of the superweak interactions (the interactions for which
$H_{12} \neq 0$) in the system.

Let us analyze some important observations following from
(\ref{Re-h11-h22}), (\ref{test-a}) and from the rigorous result
obtained in \cite{ur02-hep-ph}. The non--vanishing of the right
hand side of the relation (\ref{test}) can not be considered as
the proof that the CPT--symmetry is violated. So, there are two
general conclusions following from (\ref{Re-h11-h22}),
(\ref{Re=0}), (\ref{test-a}) and \cite{ur02-hep-ph}. The first
one: the tests based on the relation (\ref{test}) can not be
considered as CPT--symmetry tests and this is the main conclusion
of this paper. The second one: such tests should rather be
considered as the tests for the existence of new hypothetical
(superweak (?)) interactions allowing for the first order $|
\Delta S |= 2$ transitions.

On the other hand, one should remember that the non--vanishing
right hand side of the relations (\ref{test}), (\ref{test-a}) can
be considered as the conclusive proof that new interactions
allowing for the first order $|\Delta S |= 2$ transitions $K_{0}
\rightleftharpoons {\overline K}_{0}$ exist only if another
experiment, based on other principles, definitively confirms that
the CPT--symmetry is not violated in $K_{0} - {\overline K}_{0}$
system.

Unfortunately the accuracy of the today's experiments is not
sufficient to improve the estimation (\ref{mk-mk}) to the order
required by (\ref{FL3}). This especially concerns the accuracy
required by our "more realistic estimation" for
$\frac{|\Im\,(H_{12})|}{m_{0}}$. Simply it is beyond today's
experiments reach. In the light of the above estimations, keeping
in mind (\ref{Re-h11-h22}), only much more accurate tests based on
the relation (\ref{test}) can give the answer whether the
superweak interactions exists or not.

Last remark, other results \cite{ur93,ur95,ur98} obtained  within
the approximation described in Sec. 2  suggest  also that the form
of other parameters usually used to describe properties of $K_{0}
- {\overline K}_{0}$ system is different for the case $H_{12} \neq
0$ and for the case $H_{12} = 0$. This can be used as the basis
for designing other tests for the hypothetical superweak
interactions.

\end{document}